\documentstyle[epsfig]{l-aa}

\def\ltsima{$\; \buildrel < \over \sim \;$}
\def\simlt{\lower.5ex\hbox{\ltsima}}
\def\gtsima{$\; \buildrel > \over \sim \;$}
\def\simgt{\lower.5ex\hbox{\gtsima}}

\begin{document}

\def\ltsima{$\; \buildrel < \over \sim \;$}
\def\simlt{\lower.5ex\hbox{\ltsima}}
   \thesaurus{ 02(12.04.2;   
                 13.25.3;   
                 11.14.1;   
                 11.12.2)}   

\title{The X-ray background and the evolution of AGN}
\author{
Fulvio Pompilio, Fabio La Franca \& Giorgio Matt}
\institute{Dipartimento di Fisica, Universit\`a degli Studi ``Roma Tre", 
              Via della Vasca Navale 84, I--00146 Roma, Italy}
\date{Received 30 July 1999/ Accepted .. .. ..}

\maketitle

\markboth{F.Pompilio et al.:
The X-ray background and the evolution of AGN}{}

\begin{abstract}

We discuss the constraints on the AGN evolution from the cosmic X--ray
background and source counts. A
synthesis model to fit the X--ray background is presented. The
data used are a compilation from Gruber (1992), mainly based on HEAO-1
data. The spectrum of type 2 AGN has been modeled including Compton
down--scattering within the absorbing material. 
We found an improvement when a dependence on redshift
of the relative number of obscured sources is introduced. In particular,
we found a decrease of the fraction of type 2 AGN at redshifts larger
than $\sim$2.

However, only the soft X--ray counts are well reproduced, while 
the hard (5--10 keV) BeppoSAX/HELLAS source counts are underestimated
by a factor of $\sim$1.5. A satisfactory global solution, including the
good reproduction of the hard counts, is instead obtained 
if the XRB normalization, recently obtained from BeppoSAX/MECS measurements 
(about 30\% higher than HEAO-1), is adopted.


\end{abstract}

\section{Introduction}

The cosmic X--ray background (XRB) above $\sim$ 1 keV 
is known to be 
produced by integrated emission of discrete sources. XRB 
synthesis models are usually based on the so-called unification scheme for 
Active Galactic Nuclei (AGN), which 
ascribes the different observational appearances to the orientation of a 
molecular torus surrounding the nucleus. The 
intersection of the line of sight with the torus determines a type 2 AGN 
(AGN2), while the direct observation of the nucleus identifies a type 1 AGN 
(AGN1) (Antonucci \& Miller 1985). An appropriate combination of the two type 
of sources can reproduce the main characteristics of high-energy XRB, i.e. 
intensity and spectral shape (Setti \& Woltjer 1989; 
Madau et al. 1994; Matt \& Fabian 1994; Celotti et al.
1995; Comastri et al. 1995). 

In this paper, the constraints on the evolution of AGN from
the XRB and the source counts are examined, and a possible variation of the
standard model (Comastri et al. 1995), in which the AGN2/AGN1 ratio changes
with redshift, is explored. We made use of source counts data in
the 0.5--2 keV band by ROSAT (Georgantopoulos et al. 1996), in the 0.3--3.5
keV band by \emph{Einstein} (EMSS; Gioia et al. 1990), and the source counts 
provided by the BeppoSAX HELLAS survey in the 5--10 keV band (Fiore et al. 
1999; Comastri et al. 1999).  The XRB fitting procedure was 
based on the Gruber (1992) compilation of data, mostly from 
HEAO-1 in the 3-50 keV range (Marshall et al. 
1980). 

Throughout this paper, a Hubble constant ${H}_{0}=50$ Km s$^{-1}$ Mpc$^{-1}$
and deceleration parameter ${q}_{0}=0.5$ have been assumed.

\section{Models for X-ray spectra and evolution of AGN}

\subsection{AGN1 and AGN2 spectra}

The local AGN spectrum is assumed to be the sum of 
AGN1 and AGN2 spectra (${F}_{1}(E)$, ${F}_{2}(E)$
respectively) weighted by the number ratio of AGN2 to AGN1, $R$:
\[{F}_{loc}(E)\propto \left[ {F}_{1}(E)+R\times {F}_{2}(E)\right] .\]

Following Comastri et al. (1995),
a double power-law with a Compton reflection component ${F}_{r}(E)$ has been 
adopted for the AGN1 spectrum:
\[{F}_{1}(E)\propto \left\{ \begin{array}{ll}
{E}^{-{\alpha }_{1}} & \mbox{$E<1.5~keV$}\\
{E}^{-{\alpha }_{2}}{e}^{-\frac{E}{{E}_{c}}}+{F}_{r}(E)
& \mbox{$E>1.5~keV$} \end{array} \right.\]
with ${\alpha }_{2}$=0.9 
(Matsuoka et al. 1990; Pounds et al. 1990; Nandra \& Pounds 1994), and
${\alpha }_{1}$=1.3. 

The steeper, low-energy ($E<1.5$ keV) spectrum represents 
the so-called soft excess. The shape and contribution of this
component is not well known, and in many sources evidence for its very
existence is lacking altogether. As a baseline, we have adopted the same
prescription as Comastri et al. (1995), but as it now appears to be rather
extreme, we have also explored the opposite case, i.e. no soft excess
at all (see Sect. 3.1).

\begin{figure}
\begin{center}
\epsfig{file=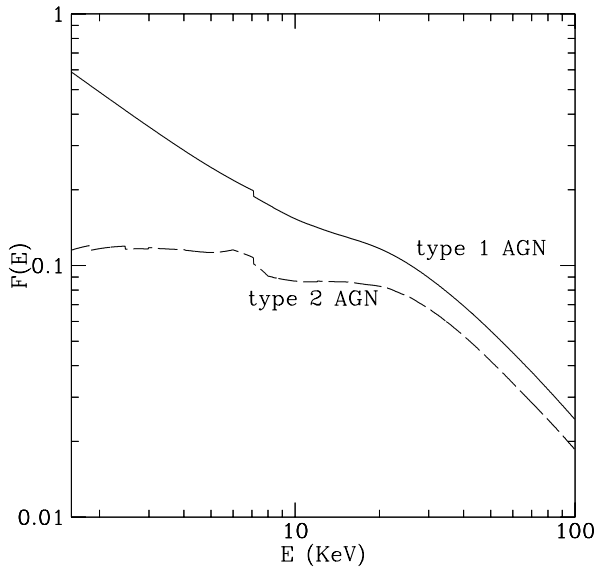,width=8cm}
\caption{The AGN1 spectrum (solid line) and the AGN2 average spectrum (dashed
line) as produced by the model.}
\end{center}
\end{figure}

The adopted value of the cut-off energy ${E}_{c}$=400 keV is also 
similar to those used in 
previous models (Comastri et al. 1995; Celotti et al. 1995) even if recent
BeppoSAX results (Matt et al. 1999a and references therein)
seem to suggest somewhat smaller values, but with a rather large spread. 
The number of sources with reliably measured values is so low, however,
that we preferred to still use the values adopted in previous models, to
make easier the comparison. In any case, 
we tested the effect of adopting lower values for the cut--off, and
found that the peak in the XRB spectrum 
at $\sim $ 30-40 keV is less well fitted.

The term ${F}_{r}(E)$ represents the Compton reflection component by 
the accretion
disk and by the torus inner surface and has been evaluated following
Magdziarz \& Zdziarski (1995), assuming an inclination angle of 60$^{\circ}$.

According to unified schemes, AGN2 spectra are obtained as 
AGN1 spectra seen through absorbing matter. The distribution
of equivalent hydrogen column density (${N}_{H}$) is chosen to be logarithmic,
i.e. $\frac{dN(log~{N}_{H})}{d(log~{N}_{H})}\propto log~{N}_{H}$,
that is a reasonable analytical approximation to the recent data on Seyfert 
galaxies (Maiolino et al. 1998; Risaliti et al. 1999).

As described in a previous paper (Matt et al. 1999b), we 
developed a transmitted spectrum model by means of Monte Carlo simulations,
assuming a spherical geometry with the X-ray source in the centre and
considering photoelectric absorption, Compton scattering and fluorescence 
(for iron atoms only), fixing element abundances as tabulated in Morrison \&
McCammon (1983). 

This transmitted component, which is relevant for  
${N}_{H}\ge {10}^{23}{cm}^{-2}$, has been so far included in XRB synthesis
models only by a handful of authors (Madau et al. 1994;
Celotti et al. 1995; Matt et al. 1999b; Wilman \& Fabian 1999).
The final spectrum ${F}_{{N}_{H}}(E)$ has been then 
averaged over the ${N}_{H}$-distribution to obtain the total AGN2 spectrum: 
\[{F}_{2}(E)=\frac{\int_{log~{N}_{H1}}^{log~{N}_{H2}}{{F}_{{N}_{H}}(E)~
log~{N}_{H}~d(log~{N}_{H})}}{\int_{log~{N}_{H1}}^{log~{N}_{H2}}{log~{N}_{H}~
d(log~{N}_{H})}}.\]

\noindent
where the 
${N}_{H}$-distribution has been considered in the range ${10}^{21}\le {N}_{H}
\le {10}^{25}{cm}^{-2}$.   

The AGN1 and AGN2 spectra are shown in Fig.~1. The AGN1 spectrum is flattened 
by the reflection component, whose contribution reaches $\sim $ 1/3 of the 
total at its maximum ($E\sim 30$ keV). The addition of a transmitted component 
significantly affects the total AGN2 
spectrum, again with a flattening  around 30-40 keV.
The Compton scattered photons in AGN2 increases the total spectrum by
$\sim $ 20\% at $\sim $ 30-40 keV with respect to a model
involving only absorption.

The last step in evaluating the overall local spectrum concerns the choice of
the number ratio $R$. Maiolino \& Rieke (1995) find $R=4.0\pm 0.9$, if
type 1.8, 1.9 and type 1.2,1.5 Seyfert galaxies are respectively classified
as AGN2 and AGN1: the estimate agrees with previous (Huchra \& Burg 1992;
Goodrich et al. 1994) and more recent (Ho et al. 1997) results, and so $R=4.0$
has been adopted.

For the sake of simplicity, iron emission line has not been included, even if
it is a common feature in AGN. However, the contribution of the line to the 
1.5-7 keV XRB is expected to be 
less then 7\% (Gilli et al. 1999a) and it is 
smeared out by the integration over the redshift range, so that XRB retains its
characteristic smoothness (Schwartz 1992), unless the emission is dominated
by a small range of redshifts (Matt \& Fabian 1994). 

\subsection{Cosmological evolution}

Hard X-rays ($3>$ keV)
are well suited for the selection of type 1 and, especially, type 2 AGNs as
they 
are less affected by absorption. Until recent past little was known about the
evolution of the AGNs in this band. The data obtained from the ASCA satellite
have allowed a first determination of the 2-10 keV AGN XLF
(Boyle et al 1997).
However the statistics was still poor. On the contrary
the AGN1 XLF in the soft X-ray band is retained to be well-known at low and
intermediate
redshift (Boyle et al. 1994; Page et al. 1996; Jones et al. 1997), while
at higher redshift ($z\ge 3$) insufficient sampling and lack of statistics
prevent the XLF from being firmly evaluated. Anyhow, X-ray AGN1 are
detected up to $z=4.6$ (Miyaji et al. 1998) and no evidence for a space 
density turn-over is found up to $z\sim 3$, likewise in the optical
(e.g. Kennefick et al. 1996) and radio surveys (Shaver et al. 1999).

For these reasons we chose to tie the AGN evolution to the
soft X-ray XLF of AGN1. We used the pure luminosity evolution
(PLE) scenario which fits the combined ROSAT and EMSS data on AGN1
space density (Boyle et al. 1994). This corresponds to
a local luminosity function that can be represented by a double power-law where
the break-luminosity ${L}_{*}$, i.e. the luminosity value corresponding to the 
slope change, evolves as ${L}_{*}(z)\propto {(1+z)}^{k}$. In the 
following, we adopt the PLE H-model of Boyle et al. (1994) (hereafter B94) in 
the 0.3-3.5 keV band:
\[\Phi ({L}_{44})=\left\{ \begin{array}{ll}
{\Phi }^{*}{L}_{44}^{-{\gamma }_{1}} & \mbox{${L}_{44}<{L}_{*}$}\\
{\Phi }^{*}{L}_{*}^{({\gamma }_{2}-{\gamma }_{1})}{L}_{44}^{-{\gamma }_{2}}
& \mbox{${L}_{44}\ge {L}_{*}$}
\end{array} \right.\]
with ${\gamma }_{1}=1.36$, ${\gamma }_{2}=3.37$, ${L}_{*}=0.39$, ${\Phi }^{*}
=1.45\times {10}^{-6}$ Mpc$^{-3}$ $({10}^{44}$ erg s$^{-1})$$^{-1}$ and 
${L}_{44}$, ${L}_{*}$ in unity of ${10}^{44}$ erg s $^{-1}$. The 
break-luminosity evolution follows:
\[{L}_{*}(z)=\left\{ \begin{array}{ll}
{L}_{*}(z=0)\times {(1+z)}^{k} & \mbox{$z<{z}_{max}$}\\
{L}_{*}(z=0)\times {(1+{z}_{max})}^{k} & \mbox{$z\ge {z}_{max}$}
\end{array} \right.\]
where $k=2.90$ and ${z}_{max}=1.73$. We limited
the analysis of AGN1 XLF to a PLE model because the most recent
attempts with pure density evolution 
models, in which the space density $\Phi (L)$ directly evolves in redshift,
overproduce the soft XRB (Hasinger 1998).
We have also to introduce an AGN2 XLF, which is a matter of strong debate. In
the framework of the unification scheme, we assumed the 
density of AGN2 to be $R$=4 times that of the corresponding unobscured AGN1.
We assumed the ${N}_{H}$-distribution 
to be independent of the AGN1 source luminosity.
A different approach may
consist in setting a completely unrelated XLF, 
but it would involve too many parameters and there are not enough
data to yield a reliable estimate. 

Boyle et al. (1997) directly measured the AGN1 and AGN2 XLF 
on a sample of 26 2--10 keV ASCA sources at a 
flux limit of $5\times {10}^{-14}$ erg cm$^{-2}$ s$^{-1}$, combined with the 
HEAO-1 AGN. The analysis 
gave a result consistent with the 0.3-3.5 keV AGN1 XLF,
albeit the evolution seems to be slower ($k=2.04$).

\section{The XRB synthesis model}

The intensity of the XRB has been calculated as follows:
\[I(E)=\int{\int{\frac{{d}^{2}N}{dVdL}\frac{{F}[L,E(1+z)]}{4\pi {d}_{l}
^{2}}dLdV}}\]
where $\frac{{d}^{2}N}{dVdL}$ is the AGN1 
XLF, ${d}_{l}$ the luminosity distance,
${F}[L,E(1+z)]$ the AGN spectrum. Introducing the comoving 
volume $dV(z)$ and the spectrum normalization $K$ corresponding to the 
considered luminosity, i.e. $K\int_{\Delta E}{{F}_{1}(E)dE}=L$
(${\Delta E}$=0.3--3.5 keV), the 
above relation can be written as:
\[I(E)=\frac{Kc}{4\pi {H}_{0}}\int{\int{L\frac{{d}^{2}N}{dVdL}\frac{{F}
_{loc}[E(1+z)]}{{(1+z)}^{2}{(1+2{q}_{0}z)}^{\frac{1}{2}}}dL}dz}\]
where the integration is performed in the range ${10}^{42}\le L\le {10}^{47}
erg~{s}^{-1}$, according to the B94 0.3--3.5 keV XLF. 
 
The XRB synthesis has been used to explore the evolution of the AGNs.
The model was fitted to the data discussed by Gruber (1992), largely
based on HEAO-1 A2 3-50 keV data\footnote{Recently, a new XRB data 
set by the UCSD/MIT Hard X-Ray and
Gamma-Ray instrument (HEAO-1 A4) has been published (Gruber et al. 1999). 
These data join smoothly with the XRB data reported by the GSFC
HEAO-1 A2 instrument, used in this paper.}. 

Best fit parameters were determined by a 
${\chi}^{2}$-minimization procedure based on the CERN
MINUIT software package.

\subsection{The baseline model}

The first analysis concerns the AGN1 high-redshift XLF. 
We tried to model
the decrease of AGN1 at high redshift by introducing a new parameter 
${z}_{d}\ge {z}_{max}$ corresponding to the onset of the space density decrease.
For redshifts in the range ${z}_{max} < z < {z}_{d}$, the density stays 
constant. We used three different shapes for  $z > z_{d}$:
{\sc i}) an exponential ($N(z)\propto {e}^{-\alpha(z-z_{d})}$); 
{\sc ii}) a polinomial ($N(z)\propto ({1+z \over 1+z_{d}})^{-\alpha}$); 
{\sc iii}) a sharp cut--off  ($N(z)=0$).  We allowed $\alpha$ to vary,
and fixed  ${z}_{d}$=4.5 (which is a simple but still good representation of 
the space density of AGN1 according to the present, sparse observations).
For any of the laws described above, the XRB is overproduced. 
We then allowed also ${z}_{d}$ to vary, always obtaining 
values of ${z}_{d}<3$. This is a value much lower than 
observed as discussed in Sect.2.2, and therefore all these
models can be considered unacceptable.   

We thus decided to adopt a sharp cut--off (model {\sc iii} above)
with ${z}_{d}=4.5$ and to fit the XRB changing  a different parameter. 
We first tried to vary the number ratio $R$,
obtaining a value of $R=3.5\pm 0.1$; the
error corresponds to 68\% confidence level, following Lampton et al. (1976).
The value is still  
in agreement with the observations. 
The fit was not very satisfactory from a statistical point of view, 
$\tilde{{\chi}^{2}}=\frac{114.4}{41}$. Part of the high value of the 
$\tilde{{\chi}^{2}}$
is certainly due to the fact that we have not included 
other classes of sources, notably starburst galaxies and clusters of
galaxies, which are likely to contribute at low energies.
However, the 
fit is bad also at energies where we expect the AGN to dominate. 
The fit shows also an excess
at higher energies ($E\ge 70$ keV), probably due to the value
assumed for the intrinsic cut-off energy; however, as underlined in section 
2.1, the choice was motivated by the aim of reproducing well
the $\sim $ 30--40 keV peak.
Even if we know that a really good fit is impossible with any smooth model, 
because of the significantly large fluctuations of the data (mostly due
to the fact that measurements from different instruments have been 
used simultaneously), 
we decided to try to improve the quality of the fit by further changing our
baseline model. 

\begin{figure}
\begin{center}
\epsfig{file=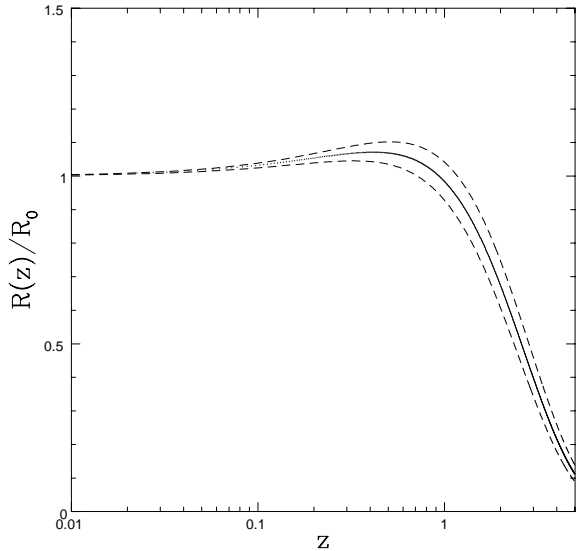,width=8cm}
\end{center}
\caption{The number ratio as a function of redshift $R(z)$ in unity of the
local value ${R}_{0}$, fixing ${R}_{0}=4.0$ (solid line). The dashed lines 
enclose the 68\% confidence region.}
\end{figure}

\begin{figure}
\begin{center}
\epsfig{file=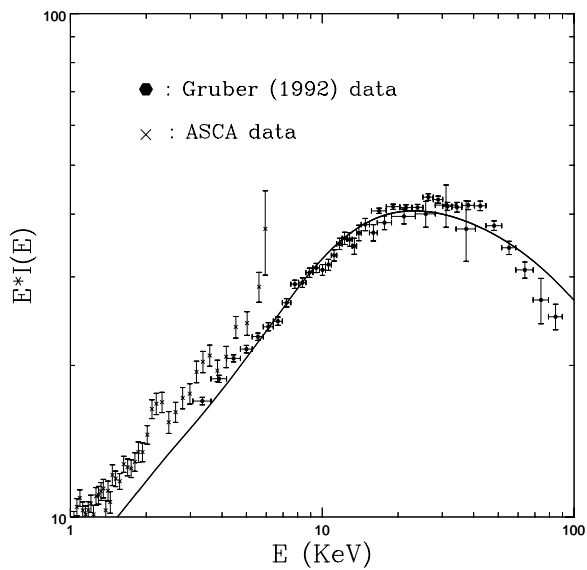,width=8cm}
\end{center}
\caption{The fit to the XRB data (solid line) with the $R(z)$ model; 
ASCA data are
also reported, even though they have not been considered in the fit.}
\end{figure}

\subsection{The R(z) model}
Because the evolution of type 2 AGNs is probably the least known 
ingredient of the model, we tried to introduce a z--dependence of the
AGN2/AGN1 ratio, $R$. 
Different analytical models
have been tried for $R(z)$: monotonic (both exponential and
power law) shapes and King profile ($R(z)\propto {\left( 1+\frac{z}{{z}_{0}}
\right)}^{-\alpha}$), but none of them gave a $\tilde{\chi}^2$ less than
$\frac{100}{41}$. 

We found, instead, a better result when a two parameter function of the 
form:
\[R(z)={R}_{0}\times {(1+z)}^{{\alpha}_{1}}{e}^{{\alpha}_{2}z}\]
is used, 
with a significant improvement of  the fit ($\tilde{{\chi}^{2}}=
\frac{91.4}{40}$) (see Figs. 2 and 3).
According to the F--test, this corresponds to a 95\%
confidence level with respect to the $R_0$=constant fit described above.
The best--fit parameters are:
${\alpha}_{1}=1.28\pm 0.04$, ${\alpha}_{2}=-0.89\pm 0.03$.
The decrease of the $\tilde{{\chi}^{2}}$ is mainly due to a better
reproduction of the data in the 5-10 keV energy range and around
the XRB energy density peak ($\sim$30 keV).  

We then checked how much dependent the above result is on the adopted
shape of the local AGN spectrum, and in particular on our choice
for the soft excess which, as discussed in Sect. 2.1,
may be rather extreme. To do that, we adopted the opposite assumption,
i.e. absence of a soft excess. The result gives an even quicker decrease
of AGN2 with redshift. This is due to the fact that in absence of the soft
excess the hard power law
has an higher normalization in order to produce the same 0.5--3.5 keV
luminosity, which implies higher luminosity at harder energies. Thus,
in order not to overproduce the XRB, less AGN2 are required.

The effect of using a pure photoelectric absorption model for
AGN2, as done in some previous papers, has also been tested. In this case the 
spectrum intensity is
higher than the photoelectric absorption + Compton scattering model (Matt
et al. 1999b), and again a steeper $R(z)$ emerges.

 It is important to note that the introduction of $R(z)$ 
implies the presence of a density evolution component for AGN2.
Density evolution has been taken into account
by the ROSAT All Sky
Survey (RASS) 0.5-2 keV data analysis of Miyaji et al. (1998),
who introduce a luminosity dependent density evolution (LDDE) basically
corresponding to a sudden drop in the evolution rate at lower luminosity. 
However, the RASS sample contains both AGN1 and AGN2, so 
the authors cannot outline any differences
between the two kind of sources and estimate their evolutionary properties 
separately. 

\begin{figure}
\begin{center}
\epsfig{file=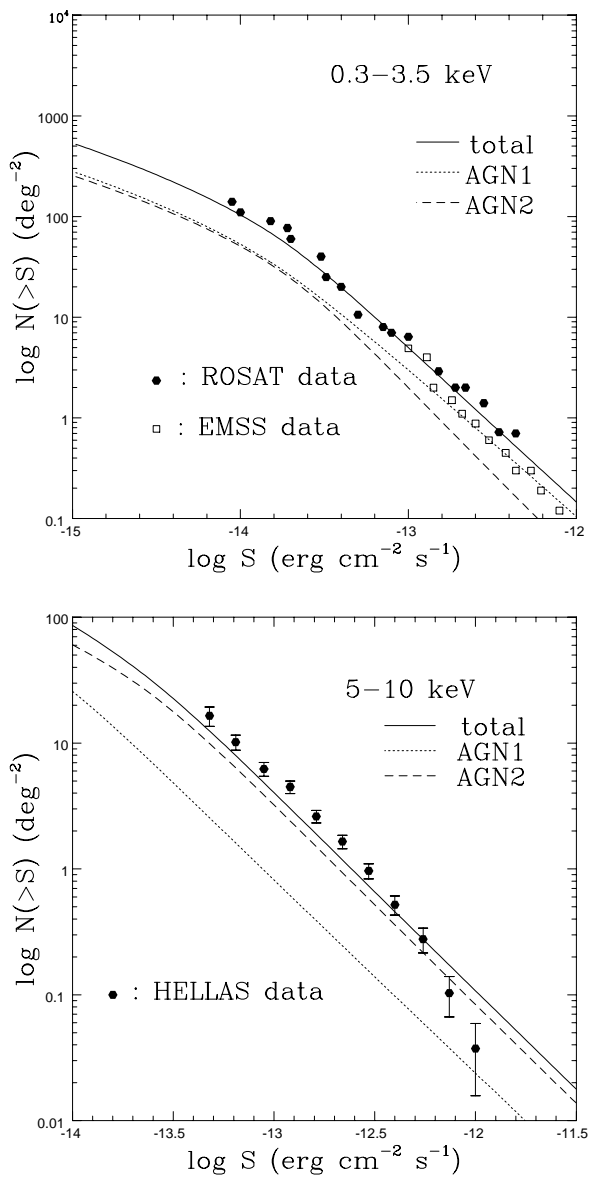,bbllx=130,bblly=34,bburx=470,bbury=710,width=8cm,
clip}
\end{center}
\caption{(Upper panel) The integral soft counts (0.3-3.5 keV) compared to 
ROSAT + EMSS data. (Lower panel) The integral hard counts (5-10 keV) compared 
to HELLAS data. The source counts have been evaluated including the 
$R(z)$ term.}
\end{figure}

So far we derived the AGN evolution only by fitting the XRB. Then we checked
the consistency of our results with the soft (ROSAT and Einstein, 0.3-3.5 keV)
and hard (BeppoSAX, 5-10 keV) counts.
In Fig.4 the source counts predicted by the model in the two bands are
compared to the ROSAT (Georgantopoulos et al. 1996) + Einstein/EMSS 
(Gioia et al. 1990) and BeppoSAX/HELLAS
(Fiore et al 1999)  data.

A good agreement in the soft band is found, while
the hard counts are underestimated by a factor of $\sim $ 1.5 at $\sim $
${10}^{-13}$ erg cm$^{-2}$ s$^{-1}$. A possible solution of this discrepancy
is discussed in the next section.

\begin{figure}
\begin{center}
\epsfig{file=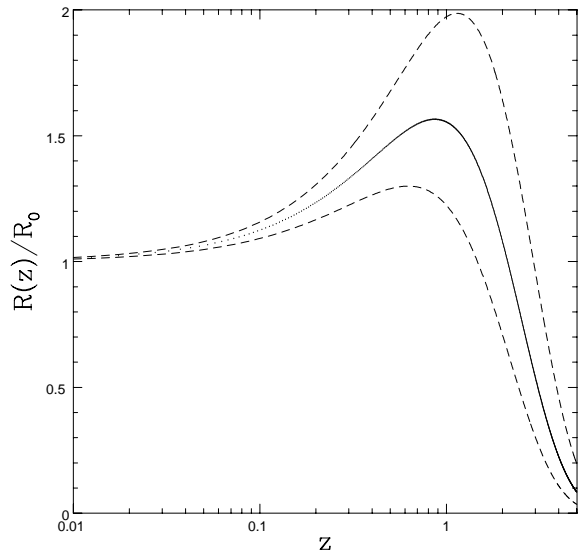,width=8cm}
\end{center}
\caption{Same as in Fig.2, but introducing a 30\% higher normalization to
XRB data.}
\end{figure}

\subsection{The normalization}

Recently, new XRB data up to 10 keV have been obtained with imaging
instruments on-board ASCA (Gendreau et al. 1995, Miyaji et al. 1998)
and BeppoSAX (Molendi et al. 1997, Parmar et al. 1999).
While there is still some disagreement between the XRB normalizations
obtained by BeppoSAX and ASCA (Parmar et al. 1999), 
both of them are higher than that obtained
by HEAO-1. For instance, BeppoSAX/MECS result is 30\% higher than
the HEAO-1 (Vecchi et al. 1999). 

So, we have fitted our model to the XRB
introducing a 30\% higher normalization to the data, but retaining
the same spectral shape.
According to the procedure discussed in Sect. 3.2, we first tried 
with a constant ${R}_{0}$  obtaining ${R}_{0}=4.9\pm 
0.1$ ($\tilde{{\chi }^{2}}=\frac{186.7}{41}$).
Again a better fit is obtained with a redshift dependence of ${R}$.
The fit to $R(z)$ 
yields ${\alpha}_{1}=2.8\pm 0.2$, ${\alpha}_{2}=-1.5\pm 0.1$ 
($\tilde{{\chi }^{2}}=
\frac{161.9}{40}$). $R(z)$ is shown in Fig.5, while the corresponding
fit is shown in Fig.6. With respect to the fit with the old normalization, 
there is a more pronounced 
increase of the fraction of AGN2 for $z$ between 0.5 and 2. 

\begin{figure}
\begin{center}
\epsfig{file=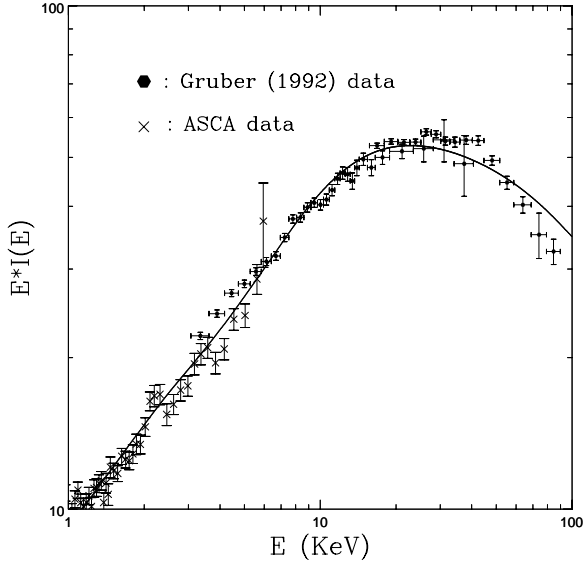,width=8cm}
\end{center}
\caption{Same as in Fig.3, but introducing a 30\% higher normalization to
the Gruber (1992) data.}
\end{figure}
\begin{figure}
\begin{center}
\epsfig{file=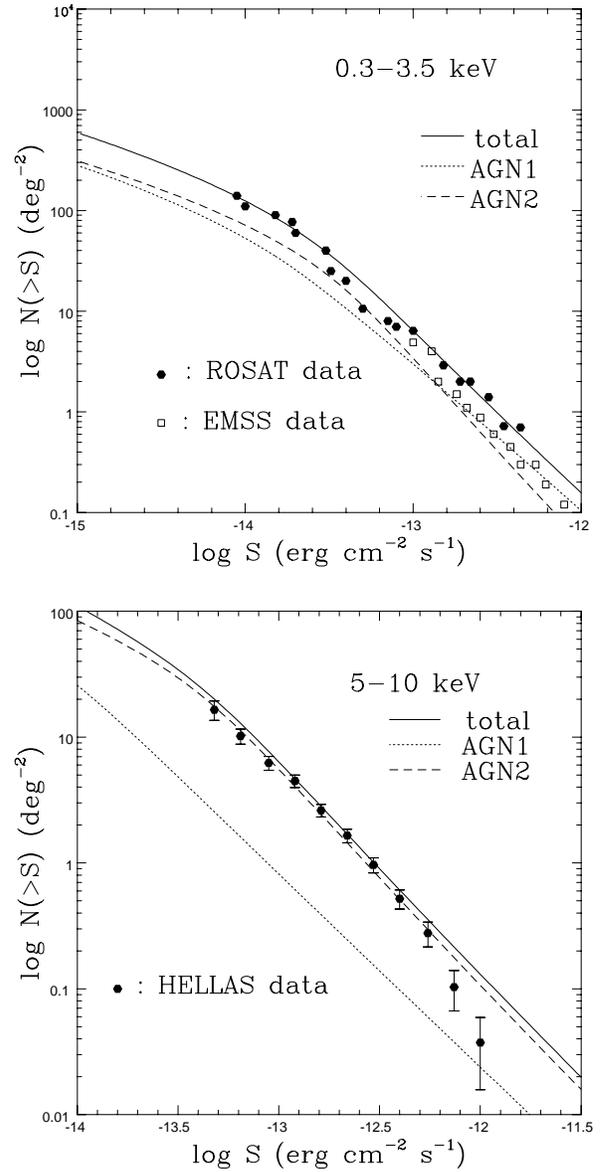,bbllx=130,bblly=34,bburx=470,bbury=710,width=8cm,
clip}
\end{center}
\caption{Same as in Fig.5, but introducing a 30\% higher normalization to
the Gruber (1992) data.}
\end{figure}

The improvement is significant at the 98\% confidence level, according to
the F-test. It is worth noting than now 
ASCA data are well reproduced, even if not used in the fit. 
This supports the higher normalization hypothesis. 
Both $\tilde{{\chi }^{2}}$ are significantly higher than the fits obtained 
with the old normalization, mainly due to a 
clear deficiency of the model below $\sim$5 keV (where, however, there may
be a contribution from other classes of sources). In this range the
contribution of the AGN1, which cannot change in our model, is most
relevant, and  it was already deficient with the lower normalization. 
At higher energies, instead, the data are well fitted thanks to the
increased number of AGN2. However, we remind that a disagreement between
the model and the data starts to emerge for $E>70$ keV. This is
unavoidable, as mentioned in Sect. 3.1, when the cut-off energy value is fixed
in order to have the XRB energy density peak at $\sim $30-40 keV.

The most interesting result is shown in Fig.7, illustrating soft and 
hard counts. 
The higher normalization enables to reproduce both. Even if the 
normalization problem must still be considered an open issue, it cannot
help being noted that with the high normalization a global solution is found. 

\section{Discussion and conclusions}

In the previous section we have discussed the possibility of a 
dependence on redshift of the AGN2/AGN1 ratio. In particular, this
ratio remains constant or slightly increases up to $z\sim$2, and then
decreases. This would suggest that AGN2 are a later
evolutionary stage of the AGN phenomenon, a possibility worth to
be explored theoretically. 

Another possibility is that the decreasing fraction of AGN2 for z$\simgt$2 is 
only apparent, and that in reality there is an increase of the fraction of
sources with $N_H\simgt10^{25}$ cm$^{-2}$, i.e. completely hidden
at all X--ray energies. This could be linked with the star formation rate
history, which is observed to increase with the redshift up to z$\sim$2,
and than stays constant (Madau et al. 1996; Rowan-Robinson 1999). 
A high star formation
rate would imply a large amount of dust and gas, and then a large
absorption.  

\begin{table}
\begin{center}
\begin{tabular}{||l|r|r|c| |}  \hline
Flux                    &  AGN1      &  AGN2      & AGN2 percentage\\
\hline
$erg~{cm}^{-2}{s}^{-1}$ &${deg}^{-2}$&${deg}^{-2}$&\%\\
\hline
${10}^{-15}$            & 162.8     & 242.1     &60\\
$5\times {10}^{-15}$    & 42.6     & 85.7    &67\\
${10}^{-14}$            & 18.3     & 48.3     &73\\
$5\times {10}^{-14}$    & 1.6     &  6.4     &80\\
\hline
\end{tabular}
\end{center}
\caption{AGN2 percentage prediction as a function of the sampling flux in the
5-10 keV band. The second and third column indicate the number density of AGN1
and AGN2.}
\end{table}
\begin{table}
\begin{center}
\begin{tabular}{||l|r|r|c||}  \hline
Flux                    &  AGN1      &  AGN2      & AGN2 percentage\\
\hline
$erg~{cm}^{-2}{s}^{-1}$ &${deg}^{-2}$&${deg}^{-2}$&\%\\
\hline
${10}^{-15}$            & 162.8     & 299.3     &65\\
$5\times {10}^{-15}$    & 42.6     & 115.2     &73\\
${10}^{-14}$            & 18.3     & 69.2     &79\\
$5\times {10}^{-14}$    & 1.6     & 11.3     &88\\
\hline
\end{tabular}
\end{center}
\caption{Same as in tab.1, but introducing a 30\% higher normalization to
the Gruber (1992) data.}
\end{table}

A different approach in fitting the XRB 
consists in a luminosity--dependent number ratio $R(L)$, 
as has already been done by Gilli et al. (1999b). 

The direct way to discriminate between different evolutionary models
is to study the AGN2 XLF, a task within the capabilities of the new generation
X--ray missions ({\it Chandra} and XMM). 


In Tab.1 and Tab.2 we report the AGN1 and AGN2 densities for different 
flux limits corresponding to the $R(z)$ models without and with the inclusion
of the 30\% increase in the normalization of the XRB. 
The effect of $R(z)$ shows up in an AGN2 percentage decreasing at lower 
fluxes, a consequence of the sampling at higher redshifts
where the  AGN2/AGN1 number ratio decreases. It is worth noting that
 ${10}^{-15}$ erg cm$^{-2}$ s$^{-1}$ and 
$5\times {10}^{-15}$ erg cm$^{-2}$ s$^{-1}$ are the flux limits expected for
the Deep observations of the Lockman Hole and the Hubble Deep Field 
scheduled for {\it Chandra} and XMM, respectively. 

\section*{Acknowledgements}

We wish to thank the BeppoSAX Scientific Data Center for assistance and the
HELLAS group for the useful and stimulating collaboration. We would also like
to thank F.Fiore, P.Giommi, A.Comastri, S.Molendi and A.Vecchi for supplying
us the data involved in this work and for several useful discussions, 
R.Gilli, G.Risaliti and M.Salvati for valuable comments on the
manuscript and the referee Dr.W.Brinkmann for suggestions to improve the 
final version.
This work was partially
supported by the Italian Space Agency, and by the Ministry for University and
Research (MURST) under grant {\sc cofin}98--02--3

\end{document}